\documentclass{article}
\usepackage{spconf,amsmath,epsfig}
\usepackage{ mathrsfs }
\usepackage{ amssymb }
\usepackage{spconf,amsmath,graphicx}
\usepackage{multirow}
\usepackage{cite}
\usepackage{amsmath,amssymb,amsfonts}
\usepackage{algorithmic}
\usepackage{graphicx}
\usepackage{textcomp}
\usepackage{xcolor}
\usepackage{xspace}
\usepackage{float}
\usepackage{amssymb, amsmath}
\usepackage[T1]{fontenc}
\usepackage{cite}
\usepackage{orcidlink}
\usepackage{academicons}
\usepackage{tabularx}
\usepackage{booktabs}
\usepackage{adjustbox}
\usepackage{nicematrix}
\usepackage{graphics}
\let\OLDthebibliography\thebibliography
\renewcommand\thebibliography[1]{
  \OLDthebibliography{#1}
  \setlength{\parskip}{0pt}
  \setlength{\itemsep}{0pt plus 0.3ex}
}

\begin{document}\sloppy

\def\x{{\mathbf x}}
\def\L{{\cal L}}

\title{KDAS: Knowledge Distillation via Attention Supervision Framework for Polyp Segmentation}
%
\name{Quoc-Huy Trinh$^{1,2, 3}$, Minh-Van Nguyen$^{1,2, 3}$, Phuoc-Thao Vo Thi$^{1,2}$}
\address{$^1$ University of Science, VNU-HCM 
$^2$ Vietnam National University, Ho Chi Minh City, Vietnam\\
$^3$ SpexAI GmbH, Dresden Germany
}

\maketitle

\begin{abstract}
Polyp segmentation, a contentious issue in medical imaging, has seen numerous proposed methods aimed at improving the quality of segmented masks. While current state-of-the-art techniques yield impressive results, the size and computational cost of these models create challenges for practical industry applications. To address this challenge, we present KDAS, a Knowledge Distillation framework that incorporates attention supervision, and our proposed Symmetrical Guiding Module. This framework is designed to facilitate a compact student model with fewer parameters, allowing it to learn the strengths of the teacher model and mitigate the inconsistency between teacher features and student features, a common challenge in Knowledge Distillation, via the Symmetrical Guiding Module. Through extensive experiments, our compact models demonstrate their strength by achieving competitive results with state-of-the-art methods, offering a promising approach to creating compact models with high accuracy for polyp segmentation and in the medical imaging field. The implementation is available on https://github.com/huyquoctrinh/KDAS.
\end{abstract}
\section{Introduction}
\vspace{-3mm}
\label{sec:intro}

Colorectal Cancer (CRC) stands as one of the most perilous diseases worldwide, emerging as a prevalent affliction affecting approximately one-third of the global population. To help the efficient treatment, the early diagnosis of CRC and Polyp segmentation tools are proposed to support the early treatment while this tool can help to localize the polyp region.

Lately, several deep learning methods have been proposed to address the challenge of polyp segmentation. In 2015, UNet \cite{unet} was introduced as an efficient approach for early-stage polyp segmentation. Subsequent methods, including ResUNet \cite{resunet}, ResUNet++ \cite{jha2019resunet++}, UNet++ \cite{unet++}, and DDANet \cite{dda}, have presented enhanced encoder architectures to improve the segmentation of polyps in endoscopic images. PraNet \cite{pranet}, a notable method, utilizes supervision techniques to refine the segmented polyp mask. Following this, several supervision-based methods \cite{pranet, metapolyp, polyp2seg, pvt, bui2024meganet} have been proposed, yielding positive results. Additionally, in recent years, various lightweight methods, such as ColonSegNet \cite{colonsegnet}, TransNetR \cite{TransNetR}, and MMFIL-Net \cite{mmfilnet}, have been introduced to create smaller models while maintaining high accuracy in predictions. Despite the promising results demonstrated by state-of-the-art methods, two main challenges persist: \textbf{(1)} the computational cost associated with heavy models and \textbf{(2)} the lack of features and generalization issues in existing lightweight models. Addressing these challenges is crucial for bridging the gap between the accuracy of sophisticated models and the practical size requirements necessary for real-world applications.

To tackle this challenge, we present the \textbf{KDAS}, a \textbf{K}nowledge \textbf{D}istillation Framework via \textbf{A}ttention \textbf{S}upervision for Polyp Segmentation. KDAS is a framework that utilizes learning through attention maps in the supervision branches to facilitate the transfer of knowledge from the larger model to the smaller model, and help the smaller model learn the strength of the larger model. Furthermore, addressing the inconsistency between student features and teacher features, as highlighted by \textit{He et al.} \cite{kdaeproblem}, is crucial to avoiding inaccuracies in segmentation results. This observation motivates us to propose the \textbf{S}ymmetrical \textbf{G}uiding \textbf{M}odules \textbf{(SGM)} to guide the model in learning from lower scales. By employing KDAS and SGM, the smaller model can assimilate knowledge from the larger model, retain the strengths of the larger model, and, importantly, mitigate the inconsistency between teacher features and student features. Consequently, this approach enhances full feature capture and improves the generalization of the smaller model, effectively addressing challenges associated with the computational cost of large models and the limitations of existing lightweight models.

In summary, our contributions encompass three key aspects:
\begin{itemize}
    \item We introduce KDAS, a novel Knowledge Distillation framework for polyp segmentation, leveraging the Attention mechanism and the Supervision concept to help the model learn the strength of the teacher model.
    \item We propose the Symmetrical Guiding Modules (SGM), designed to facilitate knowledge transfer via covariance matrices. This process creates relationships between features of the student and teacher, thereby mitigating the inconsistency between teacher features and student features.
    \item We conduct extensive experiments to assess the performance and efficiency of our methods on diverse datasets, including  ClinicDB\cite{clinicdb}, ColonDB\cite{colondb}, Kvasir-SEG\cite{kvasir-seg}, and ETIS \cite{etis}.
\end{itemize}

This paper is organized as follows: in Section~\ref{sec:related}, we briefly review existing methods related to this research. Then we propose our methods in Section~\ref{sec:method}. Experiments setup are in Section~\ref{sec:exp}. Results of the experiment and the discussion are in Section~\ref{sec:result}. Finally, we present the conclusion in Section~\ref{sec:conclusion}.

\section{Related Work}
\vspace{-3mm}
\label{sec:related}
\subsection{Polyp Segmentation}

Polyp segmentation is a task that aims to segment polyp objects from endoscopic images, thereby assisting doctors in making more accurate decisions during early diagnosis. The initial method addresses this challenge is UNet \cite{unet}, which employs an Encoder-Decoder architecture for segmentation. Subsequently, various methods based on the UNet architecture, such as UNet++\cite{unet++}, and PEFNet \cite{pefnet}, have emerged, focusing on improving feature extraction from the encoder and utilizing skip connections to address the limitations of boundary delineation in UNet. Additionally, PraNet \cite{pranet} introduces Supervision Learning as a novel approach to mitigate the sharpness boundary gap between a polyp and its surrounding mucosa. Building on this concept, subsequent methods, including MSNet \cite{MSNet}, have been proposed to further enhance the drawbacks associated with redundant feature generation at different levels of supervision. HarDNet-CPS \cite{hardnetcps} has been introduced to specifically concentrate on the lesion area. Recently, Polyp-PVT \cite{pvt} has been proposed to help suppress noises in the features and improve expressive capabilities, leading to significant results in polyp segmentation. While state-of-the-art methods have demonstrated impressive results, the challenge remains in implementing these approaches in products due to the heavy computational demands of large models. In response to this issue, several methods have been proposed. For instance, ColonSegNet \cite{colonsegnet} introduces a lightweight architecture to mitigate the trade-off between prediction accuracy and model size. TransNetR \cite{TransNetR} are Transformer-based architectures that incorporate outlier detection methods, enhancing the generalization of lightweight models. Additionally, MMFIL-Net \cite{mmfilnet} proposes a lightweight model with integrated modules to address issues related to varying feature sizes in lightweight models. 

While these methods show promising initial results, they still exhibit a significant performance gap compared to state-of-the-art models. To address this challenge, we propose the KDAS framework to transfer strength from the state-of-the-art model to student models with tiny backbones. This knowledge distillation allows the distilled model to retain the strengths of the original model while being more computationally efficient.

\subsection{Knowledge Distillation}

Knowledge Distillation is a method designed to enable a student model to learn from a teacher, initially introduced in 2015 by Hinton et al. \cite{hinton2015distilling}, utilizing Kullback-Leibler divergence loss, commonly known as KL divergence. In the context of medical image segmentation, various approaches employ distinct techniques. For instance, SimCVD \cite{simcvd} utilizes a contrastive loss to guide the student model, while MKD \cite{li2020towards} represents a Mutual Knowledge Distillation framework that thoroughly exploits modality-shared knowledge to enhance target-modality segmentation. Additionally, works by \textit{Xu et al.} integrate the attention mechanism with convolution activation map features through Knowledge Distillation to enhance the performance of compressed models. Despite the promising results of these methods, a key challenge is identified in the inconsistency between teacher and student features, as highlighted by \textit{He et al.}\cite{kdaeproblem}. Further exploration reveals that the discrepancy in feature representation between teacher and student models is the root cause of this issue. Consequently, this disparity may cause the model to learn redundant features, resulting in inaccurate predictions. To alleviate this problem, we propose the Symmetrical Guiding Modules (SGM), which assist the model in focusing on regions that share common feature representations between the student and teacher models.

%

\section{Method}
\begin{figure*}[ht]
    \centering
    \includegraphics[width=0.72\linewidth]{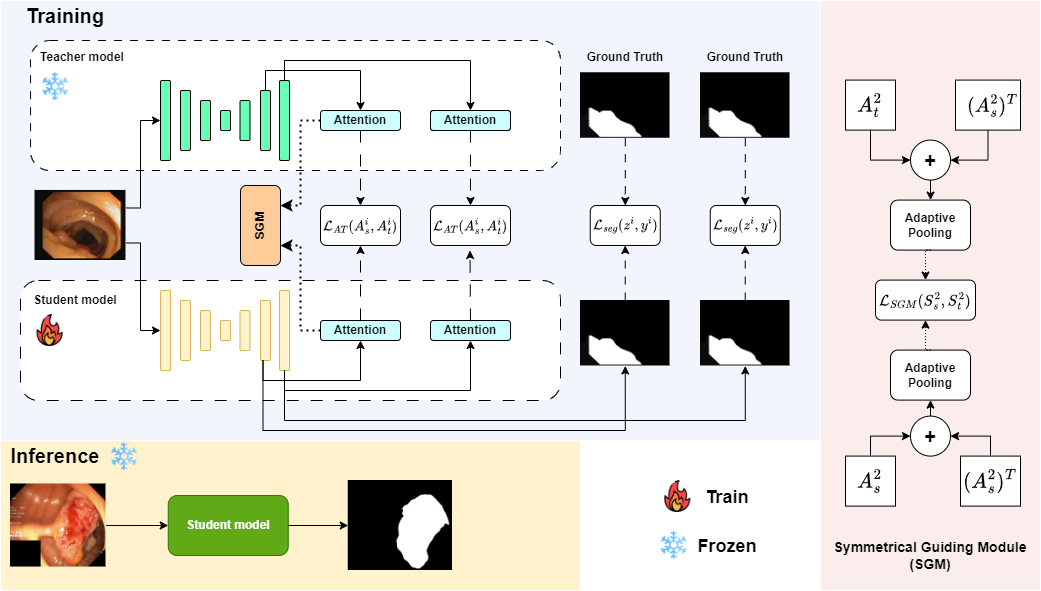}
    \caption{General KDAS framework}
    \label{fig:vachi}
    \vspace{-5mm}
\end{figure*}

\vspace{-3mm}
\label{sec:method}
In this section, we will introduce the KDAS framework, which is demonstrated in Figure~\ref{fig:vachi}. This framework includes two branches, the first branch is for the teacher model, and the other branch is for the student model. Additionally, we propose the Symmetrical Guiding Modules (SGM) to alleviate the inconsistency between the teacher features and student features. 

The input images $X$ have a shape of $B \times 352 \times 352 \times 3$, where $B$ represents the batch size. These inputs are then passed through both the teacher model $f_{t}$ and the student model $f_{s}$. Notably, the teacher model is frozen, while the student model is not frozen during the training process. The supervision outputs for each scale of both the teacher and student models, denoted as $z^{i}_{t}$ and $z^{i}_{s}$ (where $i$ indicates the order of the scale layer), are utilized for knowledge distillation.

\subsection{Attention Supervision Knowledge Distillation}

The KL divergence method enables the student model to acquire knowledge from the teacher model. However, it lacks emphasis on the critical features of the teacher's knowledge. To address this limitation, we employ the self-attention mechanism to derive attention features from the output of each scale, called Attention Supervision Knowledge Distillation (Attention$-$KD). Consequently, both the student and teacher share the same attention distribution, accentuating important information in both models. This approach effectively facilitates the student in learning crucial information from the teacher, enabling the student model to acquire the strengths of the teacher model.

The attention outputs for student features ($A^{i}_{s}$) and teacher features ($A^{i}_{t}$) are generated based on the student feature input $z^{i}_{s}$ and teacher feature input $z^{i}_{t}$. This formulation is expressed in Equations~\ref{att:student} and \ref{att:teacher}.

\begin{gather}
    A^{i}_{s} = Softmax(\frac{z^{i}_{s} \cdot (z^{i}_{s})^{T}}{\sqrt{d_{k}}})z^{i}_{s}
    \label{att:student}
    \\
    A^{i}_{t} = Softmax(\frac{z^{i}_{t} \cdot (z^{i}_{t})^{T}}{\sqrt{d_{k}}})z^{i}_{t}
    \label{att:teacher}
\end{gather}

In the context of Knowledge Distillation, the Kullback-Leibler Divergence is utilized for the process, denoted as $\mathcal{L}_{KL}$. Here, $N$ represents the number of images in a batch, and $t$ denotes the temperature value (set to 2 in our implementation, the ablation study for choosing this value can be seen in \ref{ablation:temperature}). This process is illustrated in Equation~\ref{equa:4}.

\begin{equation}
    \mathcal{L_{AT}} = (\frac{1}{N} \sum_{1}^{N}\sum^{2}_{i=1}\mathcal{L}_{KL}(A_{t}^{i}, A_{s}^{i})) \times t^{2}
    \label{equa:4}
\end{equation}

Minimizing this loss function narrows the gap between the two distributions of attention features from the teacher model and the student model. This convergence helps determine the optimal weight for the student model to learn crucial features from the teacher. As a result, the student model can learn effectively the strength of the teacher model.

\subsection{Symmetrical Guiding Module (SGM)}

As pointed out by \textit{He et al.}\cite{kdaeproblem}, the challenge in Knowledge Distillation lies in the inconsistency between the teacher's features and the student's features. To address this issue, the Symmetrical Guiding Module (SGM) is proposed. The inputs to this module include the attention outputs $A^{i}{s}$ and $A{i}^{t}$ from the last scale (when $i=2$) of student and teacher models. The symmetrical structures of the student ($S_{s}^{i}$) and teacher ($S_{t}^{i}$), which are created following to Equations~\ref{equa:1} and \ref{equa:2}.

\begin{align}
    S^{2}_{s} = \sigma(A^{2}_{s} + (A^{2}_{s})^{T})
    \label{equa:1}
    \\
    S^{2}_{t} = \sigma(A^{2}_{t} + (A^{2}_{t})^{T})
    \label{equa:2}
\end{align}

Similar to Attention$-$KD, the Kullback-Leibler divergence loss, as proposed by Hinton et al. \cite{hinton2015distilling}, is implemented to optimize the parameters of the student model. This optimization aims to minimize the distance between the two distributions of the output symmetrical structure of the student and teacher. The loss function in this stage ($\mathcal{L}_{SGM}$) is illustrated in Equation~\ref{equa:3}.
\begin{equation}
    \mathcal{L}_{SGM} = (\frac{1}{N} \sum_{1}^{N}\mathcal{L}_{KL}(S^{2}_{s}, S^{2}_{t})) \times t^{2}
    \label{equa:3}
\end{equation}

By employing a Symmetrical Structure, we can represent the features of both the student and teacher in two covariance matrices. Throughout the Knowledge Distillation process, the goal is to minimize the disparity between these two covariance matrices. This approach facilitates the establishment of relationships between the features of the teacher and the student, thereby leveraging the inconsistencies between them.

\subsection{Objective Function}

The overall Objective Function for this framework is the sum of the $\mathcal{L}_{AT}$, $\mathcal{L}_{SGM}$ and the segmentation loss ($\mathcal{L}_{seg}$ on ground truth ($z^{i}$), which is the combination of binary cross-entropy loss ($\mathcal{L}_{BCE}$) and the dice loss ($\mathcal{L}_{dice}$). The objective functions for $\mathcal{L}_{seg}$, and total loss $\mathcal{L}_{KDAS}$ is defined in Equations~\ref{equa:lseg},\ref{equa:lkdas}.

\begin{gather}
    \mathcal{L}_{seg} =  \sum^{2}_{i}\mathcal{L}_{BCE}(z_{s}^{i}, z^{i}) +  \mathcal{L}_{dice}(z_{s}^{i}, z^{i}) 
    \label{equa:lseg}
    \\
    \mathcal{L}_{KDAS} =  0.2 \times (\mathcal{L}_{AT}+ \mathcal{L}_{SGM}) + 0.8 \times \mathcal{L}_{seg}
    \label{equa:lkdas}
\end{gather}

\noindent In Equation~\ref{equa:lkdas}, 0.2 is the value that indicates the contribution of the guidance from the teacher model to the student model.
\section{Experiments}
\label{sec:exp}
\subsection{Datasets}
To conduct the fair comparison, the experiment's dataset follows the merged dataset from the PraNet \cite{pranet} experiment for the training stage which includes 900 samples from Kvasir-SEG \cite{kvasir-seg} and 550 samples from CVC-ClinicDB \cite{clinicdb}. The remaining images of Kvasir-SEG \cite{kvasir-seg} and CVC-ClinicDB \cite{clinicdb} with three unseen datasets such as ColonDB \cite{colondb}, CVC-300 \cite{cvc300}  and ETIS \cite{etis} are used for benchmarking our method.

\subsection{Implementation Detail}

In our implementation, we conducted experiments based on the Polyp-PVT baseline \cite{pvt}. We utilized the PyTorch framework and employed a Tesla V100 32GB for training. The images were resized to $352 \times 352$, and a batch size of 16 was set. For the segmentation loss ($\mathcal{L}_{seg}$), we used the Jaccard loss function, as mentioned in \cite{jaacaard}. The AdamW optimizer was employed with a learning rate of $1e-4$. The best weights were obtained after 120 epochs, with a total training time of approximately 2 hours. In the training setup, the temperature value $t$ was set to 2. During testing, the images were resized to $352 \times 352$.  

\subsection{Evaluation Metrics}

Two commonly used evaluation metrics, mean Dice (mDice) and mean Intersection over Union (mIoU), are employed for assessment. Higher values for both mDice and mIoU indicate better performance. In the context of polyp segmentation, the mDice metric holds particular significance as it is considered the most important metric for determining the effectiveness of a model.




\section{Result}
\vspace{-1mm}
\label{sec:result}
\subsection{Performance Comparisons}

\textbf{Comparison with State-of-the-art} To assess the effectiveness of our model, we compare our PVTV0 distilled model (approximately 3.7 million parameters) with several methods that have a higher parameter count. These include UNet \cite{unet}, UNet++ \cite{unet++},  PraNet \cite{pranet}, MSNet \cite{MSNet}, HarDNet-CPS \cite{hardnetcps} (2023), and PEFNet \cite{pefnet}. In this comparison, we also provide information about the number of parameters to evaluate the impact of model size on overall performance. As the datasets used in PEFNet \cite{pefnet} differ, we retrain both methods on the same settings as PraNet \cite{pranet} for a fair and consistent comparison.

\textbf{Comparison with real-time methods:}  To evaluate the performance of our compact model in terms of both prediction accuracy and computational efficiency, we conducted comparisons with several methods, namely ColonSegNet\cite{colonsegnet},  TransNetR\cite{TransNetR}, and MMFIL-Net\cite{mmfilnet}. Except for MMFIL-Net, we reproduced the training processes for the remaining models using the same training dataset as our models. The comparison weights were carefully selected to ensure a fair and meaningful comparison.


\subsection{Qualitative Results}

\textbf{Comparison with State-of-the-art:} As depicted in Table~\ref{table:quality}, our Knowledge Distillation framework significantly enhances the performance of the compact model, particularly on the Etis and ColonDB datasets. While the results on the ClinicDB and Kvasir Dataset are slightly lower, they still remain competitive with previous methods. These findings underscore the generalization ability of our tiny model in both familiar and unfamiliar domains, despite the substantially lower number of parameters in the distilled model compared to existing methods. This observation emphasizes the promising results of our contribution in creating the compact model, demonstrating its accuracy in predictions similar to that of larger models.

\begin{table*}[htbp]
\centering

\label{tab:ablation}
\begin{tabular}{@{}lcccccccccc@{}}
\toprule
\multirow{2}{*}{Method} & \multirow{2}{*}{Params(M)} & \multicolumn{2}{c}{ClinicDB}    & \multicolumn{2}{c}{ColonDB}     & \multicolumn{2}{c}{Kvasir}      & \multicolumn{2}{c}{ETIS}               \\
                              &      & mDice          & mIoU           & mDice          & mIoU           & mDice          & mIoU           & mDice     & mIoU            \\ \midrule
UNet (2015)  \cite{unet}     &   7.6     & 0.824 & 0.767      & 0.519     & 0.449                           & 0.821        & 0.756  & 0.406         & 0.343                    \\
UNet++ (2018)   \cite{unet++} &  9.0   & 0.794  & 0.729         & 0.483          & 0.410             & 0.820       & 0.743             & 0.401      & 0.344                          \\
PraNet (2019) \cite{pranet} & 32.6 & 0.899 & 0.849 & 0.712  & 0.640 & 0.898   & 0.840 & 0.628  & 0.567 
\\
MSNet (2021) \cite{MSNet} & 29.7 & \underline{0.921} & \underline{0.879} & \underline{0.755}  & \underline{0.678} & 0.907 & \textbf{0.862} & \underline{0.719} & \underline{0.664}  
\\
PEFNet (2023) \cite{pefnet} & 28.0 & 0.866 & 0.814 & 0.710  & 0.638 & 0.892 & 0.833 & 0.636 & 0.572 
\\
HarDNet-CPS (2023) \cite{hardnetcps} & $--$ & 0.917 &  \textbf{0.887}  & 0.729 & 0.658 & \underline{0.911}  & \underline{0.856} &  0.69 & 0.619 
\\
\hline
\textbf{KDAS} & \textbf{3.7} & \textbf{0.925} & 0.872 & \textbf{0.759} & \textbf{0.679} & \textbf{0.913}  & 0.848 & \textbf{0.755} & \textbf{0.677}  &  \\ \bottomrule
\end{tabular}
\label{table:quality}
\caption{Qualiative results distilled model from KDAS on various dataset}
\end{table*}

 \textbf{Comparison with Real-time model:} 
As depicted in Table~\ref{table:realtime}, KDAS outperforms previous real-time models across four different datasets and in all metrics. Notably, our resulting model exhibits the smallest number of parameters and the lowest number of FLOPs, underscoring the effectiveness and advancement of our method compared to previous real-time models. These results highlight how KDAS, through the incorporation of Knowledge Distillation and our proposed modules, significantly contributes significant improvement to the compact model in the field of Polyp Segmentation.

\begin{table*}[htbp]
\centering

\label{tab:ablation}
\begin{tabular}{@{}lcccccccccccc@{}}
\toprule
\multirow{2}{*}{Method} & \multirow{2}{*}{Params(M)}  & \multirow{2}{*}{FLOPs (G)}  & \multicolumn{2}{c}{ClinicDB}    & \multicolumn{2}{c}{ColonDB}     & \multicolumn{2}{c}{Kvasir}      & \multicolumn{2}{c}{ETIS}               \\
                               & &            & mDice          & mIoU           & mDice          & mIoU           & mDice          & mIoU           & mDice      & mIoU         \\ \midrule
ColonSegNEt (2021) \cite{colonsegnet} & 5.0 & 6.22 & 0.28 & 0.21 & 0.12 & 0.10 & 0.52 & 0.39  & 0.16 & 0.12\\
TransNetR (2023) \cite{TransNetR} & 27.3 & 10.09   & 0.87 & 0.82 & 0.68 & 0.61  & 0.87 & 0.80 & 0.60 & 0.53  \\
MMFIL-Net (2023) \cite{mmfilnet} & 6.7 & 4.32 & 0.890 & 0.838  & 0.744 & 0.659 & 0.909 & \textbf{0.858} & 0.743  & 0.670 \\

\hline
\textbf{KDAS}    & \textbf{3.7} &    \textbf{2.01}& \textbf{0.925} & \textbf{0.872} & \textbf{0.759} & \textbf{0.679} & \textbf{0.913}  & 0.848 & \textbf{0.755} & \textbf{0.677}\\ \bottomrule
\end{tabular}
\label{table:realtime}
\caption{Comparison of  model from KDAS with Real-time models}
\end{table*}

\subsection{Qualitative Visualization}
\label{subsec:vis}

\begin{figure}[H]
\vspace{-2mm}
    \centering
    \includegraphics[width= 0.75\linewidth]{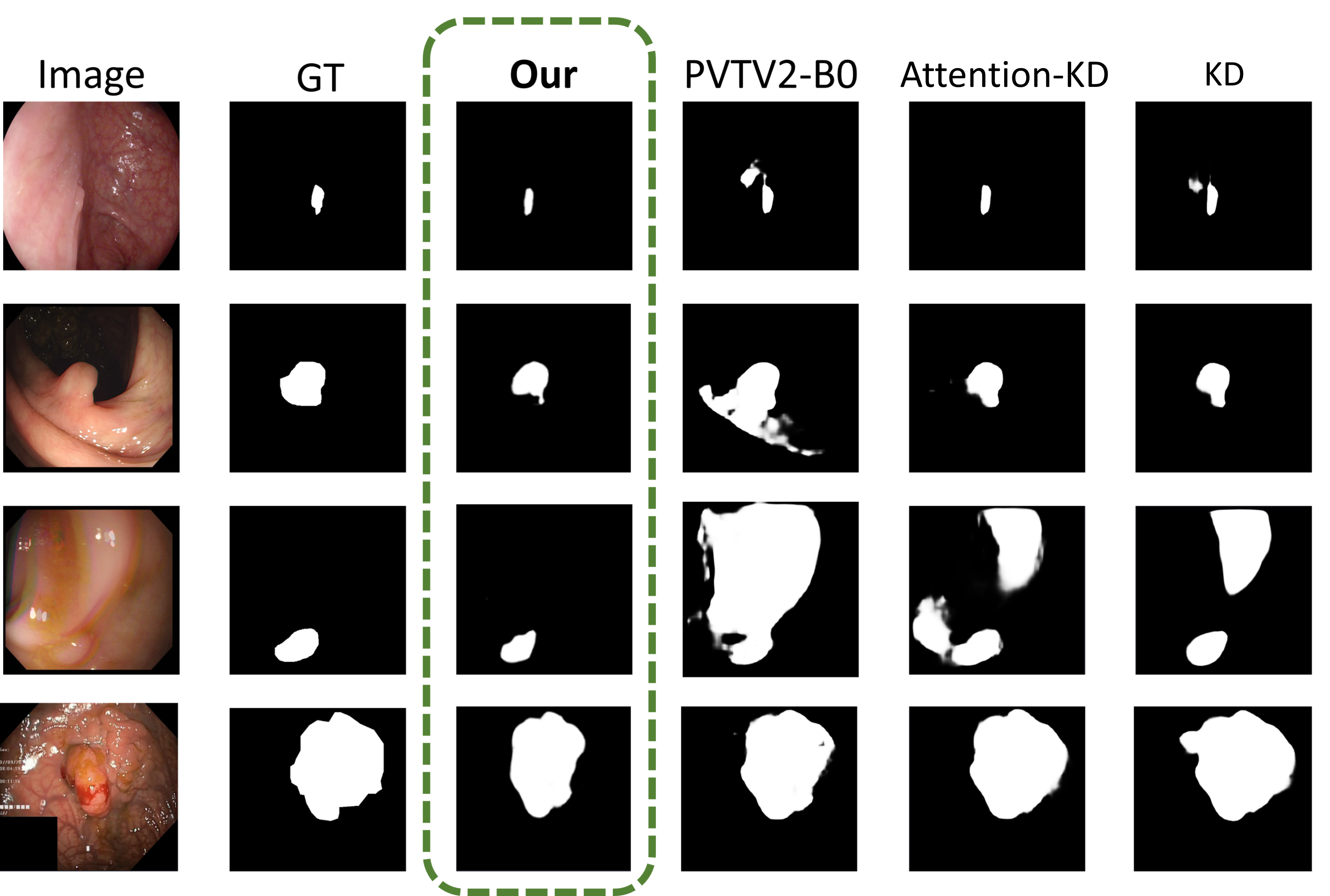}
    \caption{Qualitative results for the improvement by integrating KDAS}
    \label{fig:vis}
    
\end{figure}

In Figure~\ref{fig:vis}, we visually demonstrate the improvements achieved by integrating KDAS through qualitative visualization. The results showcase the training model progression from the baseline Polyp-PVT with B0 backbone, models with guidance from Knowledge Distillation, Attention$-$KD, to our KDAS. All visualizations are presented in Figure~\ref{fig:vis}.

From Figure~\ref{fig:vis}, it is evident that the integration of KDAS enables the model to better capture polyps overall, including tiny objects. These results highlight the improvement of KDAS in guiding the student model with small parameters to effectively segment data across various domains.

\subsection{Ablation Studies}
\subsubsection{Effect of Attention$-$KD and SGM modules}

To assess the effectiveness of the Attention$-$KD and SGM modules, quantitative experiments were conducted using the baseline Polyp-PVT with B0 backbone trained on our dataset setting. Initially, KL divergence was implemented, followed by the integration of Attention$-$KD into the baseline. Finally, the SGM, representing the full method of the KDAS framework, was added. All student models were tested on the Etis Dataset, and the results are presented in Table~\ref{tab:ablation}.

The results demonstrate that training with guidance from the teacher model enhances the performance of the student model. Moreover, the incorporation of Attention$-$KD and SGM into the Knowledge Distillation pipeline allows the student model to focus on the important information from the teacher. This approach also helps address inconsistencies between features of different scales in both the student and teacher models, leading to improved results overall.

\begin{table}[h]
\begin{center}
\begin{tabular}{|c| c | c| c|} 
 \hline
 Model & mDice & mIoU   \\ [0.5ex] 
 \hline
Baseline  & 0.717 & 0.642  \\
KL divergence & 0.747 & 0.675  \\
Attention-KD  & 0.735 & 0.649  \\
Attention-KD SGM & \textbf{0.755}& \textbf{0.677}   \\
 
 \hline
\end{tabular}
\end{center}
\caption{Ablation study for effect of Attention$-$KD and SGM modules on the Etis dataset \cite{etis}}
\label{tab:ablation}
\end{table}
\vspace{-3mm}
\subsubsection{Effect of temperature values in general framework}
\label{ablation:temperature}
To investigate the influence of temperature in the KDAS framework, we conducted experiments using different values on the ColonDB dataset and compared the results using the \textit{mDice} metric, as shown in Table~\ref{tab:ablation_t}. The rise in temperature necessitates a gradual increase in learning difficulty for students.
\begin{table}[h]
\begin{center}
\begin{tabular}{|c| c | c| c| c|c|} 
 \hline
 Temperature & 1 & \textbf{2} & 4 & 6 & 8\\ [0.5ex] 
 \hline
Dice-score & 0.756 & \textbf{0.759} & 0.755 & 0.751 & 0.748\\
\hline

\end{tabular}
\end{center}
\caption{Ablation study of temperature value on the ColonDB dataset \cite{colondb}}
\label{tab:ablation_t}
\end{table}
\vspace{-3mm}
As depicted, the model achieves its peak performance when the temperature is set to 2. However, as the temperature increases, the model's scores decline. This observation indicates that exceeding the specified temperature value could be detrimental to the student learning process.
\vspace{-3mm}
\section{Conclusion}
\label{sec:conclusion}
In conclusion, we introduce KDAS, a Knowledge Distillation Framework via Attention Supervision designed to help the model assimilate the robust knowledge from the teacher, resulting in a more efficient and compact model. Additionally, our approach incorporates a Symmetrical Guiding Module to address the inconsistency between student and teacher features. Our method demonstrates effectiveness in enabling the student model to acquire the strengths of the teacher model. The student model not only achieves competitive results compared to state-of-the-art methods but also outperforms previous real-time methods and shows improvement over the baseline, as indicated in our ablation studies. This method holds promise for further research in compact medical segmentation models, facilitating their integration into real-world applications.
\vspace{-3mm}
\section{Acknowledgement}
\vspace{-3mm}
This research is supported by research funding from Faculty of Information Technology,
University of Science, Vietnam National University - Ho Chi Minh City.



\footnotesize
\bibliographystyle{IEEEbib}
\bibliography{icme2023template}

\end{document}